# Dramatic enhancement of spin-spin coupling and quenching of magnetic dimensionality in compressed silver difluoride†

Dominik Kurzydłowski,*[a,b] Mariana Derzsi,[a,c] Paolo Barone,[d] Adam Grzelak,[a,e] Viktor Struzhkin,[f] José Lorenzana,*[g] and Wojciech Grochala*[a]

a. Center of New Technologies, University of Warsaw, ul. Banacha 2c, 02-097 Warsaw, Poland

b. Faculty of Mathematics and Natural Sciences, Cardinal Stefan Wyszyński University, ul. Dewajtis 5, 01-038 Warsaw, Poland

c. Advanced Technologies Research Institute, Faculty of Materials Science and Technology in Trnava, Slovak University of Technology in Bratislava, J. Bottu 25, 917 24 Trnava, Slovak Republic

d. SPIN-CNR, via Vetoio, 67100 L'Aquila, Italy.

e. Faculty of Chemistry, University of Warsaw, ul. Pasteura 1, 02-093 Warsaw, Poland

f. Geophysical Laboratory, Carnegie Institution of Washington, 5251 Broad Branch Road NW, Washington, DC 20015, United States

g. ISC-CNR, Dipartimento di Fisica, Università di Roma "La Sapienza", Piazzale Aldo Moro 5, 00185 Roma, Italy

† Ján Gažo in memoriam

Electronic Supplementary Information (ESI) may be found at the end of this manuscript.

**Abstract**

Meta-GGA calculations of the ambient and high-pressure polymorphs of silver difluoride indicate that the compression-induced structural changes lead to a 3.5-fold increase in the strength of antiferromagnetic spin-spin interactions resulting in coupling constant values higher than those found for record-holding oxocuprates(II).

AgF$_2$, containing the Ag$^{2+}$ cation ($4d^9$ electronic configuration), differs in many ways from other metal fluorides. Due to the cooperative Jahn-Teller effect[1] this compound adopts a unique structure featuring puckered sheets of [AgF$_2$] stoichiometry (Fig. 1a).[2] Moreover the bonding between Ag$^{2+}$ and F$^-$ has a considerable covalent component.[3] This two features lead to strong 2D antiferromagnetic (AFM) interactions between the spins of unpaired electrons residing on neighbouring Ag$^{2+}$ centres.[2] Indeed, AgF$_2$ turns out to be an excellent analogue of layered oxocuprates, in particular La$_2$CuO$_4$.[4]

In both AgF$_2$ and La$_2$CuO$_4$ strong 2D spin-spin (magnetic) interactions are mediated through the superexchange (SE) mechanism, and thus strongly linked to the structural features, in particular bond lengths, and angles. Here, we presents computations,‡ based on the SCAN meta-GGA functional,[5] on the spin-spin interactions in the ambient and high-pressure polymorphs of AgF$_2$. Recent experiments conducted for AgF$_2$ at pressures up to 40 GPa (400 kbar) revealed that the ambient-pressure *Pbca* structure (denoted LP) transforms above 7 GPa to a polymorph of *Pca2$_1$* symmetry (here as HP1);[6] this is a distorted variant of LP also comprising [AgF$_2$] sheets (Fig. 1b). Above 15 GPa HP1 transforms into the HP2 polymorph (*Pbcn* symmetry) which features nanotubular units (Fig. 1c).[6,7]
All of these structural motifs share the same building block – square [AgF$_4$]$^{2-}$ units with short Ag-F bonds (2.0–2.1 Å); each square is connected with four others via a single F-atom bridge. These units are planar for LP, while for HP1 and HP2 the Ag$^{2+}$ cation resides slightly above the plane formed by the four F$^-$ anions. The square coordination of Ag$^{2+}$ forces the unpaired electron to reside in a $d(x^2-y^2)$-type orbital lying in the plane of the [AgF$_4$]$^{2-}$ units. In light of the Goodenough-Kanamori-Anderson rules established for the SE mechanism,[8] one would expect strong spin-spin coupling only between [AgF$_4$]$^{2-}$ units connected by short Ag-F bonds and bridged by a single F-atom, that is the interactions should be considerable only for SE paths within sheets (LP, HP1) or nanotubes (HP2), and much weaker for inter-sheet and inter-tube interactions. This is indeed what is found in our calculations, as will be shown below.

As shown in Fig. 1a for LP there is only one type of SE route within the [AgF$_2$] sheets; we mark it (and the corresponding SE constant) as $J_{2D}$. For the HP1 structure alternation in the values of the Ag-F-Ag angles and Ag-F bond lengths leads to two paths: $J'_{2D}$ and $J''_{2D}$ (Fig. 1b). Finally there are three distinct SE paths within the nanotubes of HP2: $J'_t$, $J''_t$, and $J'''_t$ (Fig. 1c).
In Fig. 2a we present the sum of the Ag-F bond lengths along the respective coupling paths, $R_{Ag-F-Ag}$ (structural data is taken from ref. 6). The variation in $R_{Ag-F-Ag}$ does not exceed 0.1 Å (<3%) up to pressures of 40 GPa, which is a manifestation of the rigidity of the Ag-F bond.[3,9] More pronounced changes are found for the angle of the Ag-F-Ag bridge ($\alpha_{Ag-F-Ag}$, Fig. 2b). Compression of LP leads to a slight decrease of $\alpha_{Ag-F-Ag}$, which is a manifestation of the increased puckering of the [AgF$_2$] sheets. Upon the LP-HP1 transition rotation of the [AgF$_4$]$^{2-}$ squares within the sheets leads to an alternation in $\alpha_{Ag-F-Ag}$ with the Ag-F-Ag bridge along $J'_{2D}$ becoming more linear while that along $J''_{2D}$ more bent. Above 15 GPa HP1 is found experimentally to transform into the HP2 polymorph with its three types of Ag-F-Ag bridges. Those along the $J'_t$ path are nearly linear (~175°), while bridges along $J''_t$ exhibit angles comparable to those found in LP and HP1. Finally, $\alpha_{Ag-F-Ag}$ is close to 90° for $J'''_t$. As shown both by theoretical analysis,[10,11] as well as experimental data,[12,13] changes in the angle of single bridge between two $d^9$ complexes can lead to a huge variation in the coupling constant. We therefore performed calculations of the coupling constants along the above mentioned SE pathways, as well as those characterizing weaker inter-sheet and inter-tube couplings. For this we utilized the broken-symmetry method,[14,15] as applied in ref 16 (*cf.* ESI). We used the SCAN meta-GGA functional,[5] which was found to correctly describe the electronic and magnetic properties of transition metal compounds,[17] in particular spin-spin coupling constants.[18] Importantly, SCAN was found here to well reproduce the experimental intra-sheet coupling constant ($J_{2D}$) of the LP structure of $AgF_2$ at ambient conditions (theor.: −71 meV; exp.: −70 meV).[4]

Our results confirm the 2D AFM nature of the LP structure at ambient and elevated pressures (Fig. 3a and ESI Fig. 1a) with the absolute value of $J_{2D}$ being an order of magnitude larger than the inter-sheet interactions (Fig. 2c).§ Upon compression the intra-sheet AFM interaction becomes weaker ($J_{2D}$ changes from −71 meV at 0 GPa to −55 meV at 7 GPa) due to the bending of the Ag-F-Ag bridge (Fig. 2b). This contrasts the high-pressure behaviour of the layered oxocuprates(II) where compression leads to an increase of the 2D AFM interactions due to bond compression at no additional layer puckering.[19–21]

The alternation in the Ag-F-Ag bridge angle upon the LP-HP1 transition at 7 GPa leads to two different values of the coupling constants with the more negative one ($J'_{2D}$) associated with a SE route via the more linear Ag-F-Ag bridge. Upon compression $\alpha_{Ag-F-Ag}$ along this route exhibits little change and consequently $J'_{2D}$ remains nearly constant at *ca.* −75 meV. In contrast the pressure-induced decrease of $\alpha_{Ag-F-Ag}$ for the $J''_{2D}$ SE route leads to considerable weakening of the AFM interaction in this direction. In fact at 15 GPa $J''_{2D}$ becomes slightly positive (0.5 meV) and comparable in strength to the inter-sheet interactions (Fig. 2c). At this point the structurally 2D HP1 phase can be described as magnetically 1D as $J'_{2D}$ becomes the dominant SE interaction and forms a zig-zag 1D AFM chains running along the crystallographic **b** direction (ESI, Figure 1b).

Above 15 GPa $AgF_2$ adopts the nanotubular HP2 form where the intra-tube SE routes exhibit an even more diverse range of Ag-F-Ag angles (Fig. 2c). Not surprisingly our calculations yield three very different values of the intra-tube couplings. The largest difference can be seen for $J'_t$ and $J'''_t$ – while the former exhibits extremely strong AFM coupling ($J'_t \approx$ −250 meV), the other is weakly to moderately ferromagnetic (4 meV < $J'''_t$ < 30 meV). The $J''_t$ route connects neighbouring $[AgF_4]^{2-}$ units into $[AgF_3]^-$ chains and it is moderately AFM, its value (*ca.* −25 meV) being an order of magnitude smaller than $J'_t$. The emerging picture is that despite its 1D nanotubular structure the HP2 polymorph is magnetically 0D *i.e.* comprised of AFM-coupled $[Ag_2F_7]^{3-}$ dimers. Analysing the topology of the couplings (see ESI Fig. 1c) we conclude that the magnetic ground state is most likely gaped due to singlet formation along the $J'_t$ coupled bridges.[22] Analogous dimers, but exhibiting a Ag-F-Ag angle of 180°, are found in the complex fluoride $Ag_2ZnZr_2F_{14}$.[23] Our calculations of the intra-dimer SE coupling for this compound yielded a value of −313 meV indicating that the previous analysis has probably underestimated the strength of SE coupling in this important reference system.§§ The intra-dimer AFM coupling in $Ag_2ZnZr_2F_{14}$ being stronger than that for the HP2 phase is a consequence of the linearity of the Ag-F-Ag bridge, and the shorter Ag-F bonds along it ($R_{Ag-F-Ag}$ =4.03 Å for $Ag_2ZnZr_2F_{14}$ while it ranges from 4.11 to 4.15 Å for HP2). Overall the high-pressure transitions of $AgF_2$ lead to a progressive lowering of the magnetic dimensionality from 2D (LP phase), through a 2D to 1D transition observed upon compression of the HP1 polymorph in the 7–15 GPa range, up to a dramatic collapse into a dimeric (0D) albeit structurally nanotubular magnetic structure upon the HP1-HP2 transition at 15 GPa. This is reminiscent of the reduction of magnetic dimensionality observed for the $Cu^{2+}$-containing $CuF_2(H_2O)_2$(pyrazine) upon compression,[24] or the enhancement of 1D properties predicted for CuO and $AgFBF_4$ at high pressures.[9,12,13]

The main driving force behind the magnetic dimensionality reduction are the changes in the angles of the Ag-F-Ag bridge which determine the hopping matrix elements between the different orbitals. This is most clearly seen in Fig. 3a where the $J$-values obtained for various $AgF_2$ polymorphs are plotted against $\alpha_{Ag-F-Ag}$. One can understand the main trends using perturbation theory in the hybridization parameterized through Slater-Koster[25] matrix elements. The magnetic interaction for a single bridge can be written as $J = J^{(2)} + J^{(4,SE)} + J^{(4,HR)}$ where $J^{(2)}$ is a second order ferromagnetic contribution due to direct exchange between the Ag and F, $J^{(4,SE)}$ is the proper SE contribution and $J^{(4,HR)}$ is a ferromagnetic contribution due to Hund's exchange in F (for details see ESI). Fig. 3a shows that this model indeed reproduces the main trends. The strong angle dependence of the AFM coupling strength connected with the formation of nearly linear Ag-F-Ag bridges in HP2 leads to a 3.5-fold increase in the SE interaction strength upon compression to 15 GPa. Remarkably, the intra-dimer coupling constant ($J'_t$) found here for HP2 surpasses that found for oxocuprates(II), both magnetically 2D ($La_2CuO_4$, $J_{2D}$ = −146 meV)[26] and 1D ($Sr_2CuO_3$, $J_{1D}$ = −241 meV).[27] It is also higher than those found for the majority of F-bridged magnetic systems,[28] with the exception of $AgFBF_4$ (1D system).[9]

We performed also DFT computations taking into account spin-orbit coupling and the possibility of non-collinear ground states. In excellent agreement with experiment[2] we find that for the LP form the ground state belongs to the magnetic space group $P$b'c'a. The magnetic moments are approximately directed in the ***a*** direction (see Supplementary Figure 1) and have an ordered moment of $0.6\mu_B$. The reduction from $1\mu_B$ is due to covalence effects and does not take into account zero point quantum fluctuations which should also be present. The ferromagnetic moment is $0.04\mu_B$/Ag, somewhat larger than the reported experimental value (0.01$\mu_B$/Ag) but of the correct order of magnitude. In addition to the weak ferromagnetic moment in the ***c*** direction, as expected for *P*b'c'a, there is also a weak ferromagnetic moment in the ***b*** direction (0.05$\mu_B$/Ag), however this moment is opposed on neighbouring layers and therefore does not contribute to the uniform magnetization. In the HP1 phase the ***c***-axis ferromagnetic moment remains the same but the ***b***-axis intralayer moment increases to 0.07$\mu_B$/Ag. Interestingly, this is a multiferroic phase showing a rare coexistence of ferroelectricity and ferromagnetism. Finally for the HP2 phase ferromagnetism disappears and magnetic moments align parallel to the nanotube axis.

Given the remarkable magnetic properties of the high-pressure phases of AgF$_2$ computed here, it is of interest to examine electronic properties of these phases, as well. Our hybrid functional HSE06 study shows that the fundamental band gap of AgF$_2$ (Figure 3b) is quite substantial, ~2.40–2.55 eV, and it remains nearly constant in the entire studied pressure range (0–40 GPa); the variation of the bandgap does not exceed 3 %. The robustness of the bandgap is in line with preserved connectivity of the building blocks (the square [AgF$_4$]$^{2-}$ units) and rigidity of the Ag-F bond (see Fig. 2a). Persistence of the broad band gap is typical of undoped d$^9$ systems,[21,29–32] and it points out to the robustness of the localized AFM interactions. These interactions which lead to insulating behaviour tend to act opposite to the band broadening which is usually seen at elevated pressure and consequently the band gap closing becomes much more difficult than for the diamagnetic semiconductors.

In conclusion, we have theoretically studied the recently discovered high pressure phases of AgF$_2$. The calculations suggest substantial lowering of magnetic dimensionality due to changes in Ag–F–Ag angles, with a concomitant spectacular increase of the localized magnetic interactions via F$^-$ anions, and preservation of the broad band gap even at 40 GPa.

The authors acknowledge the support from the Polish National Science Centre (NCN) within grant no. UMO-2016/23/G/ST5/04320. This research was carried out with the support of the Interdisciplinary Centre for Mathematical and Computational Modelling (ICM) University of Warsaw under grant no. GA67-13. J.L. is supported by Italian MAECI under collaborative Projects SUPERTOP-PGR04879 and AR17MO7.

**Conflicts of interest**

There are no conflicts to declare.

**Notes and references**

‡ Calculations, utilizing the SCAN functional,[5] were performed with the projector–augmented-wave (PAW) method, as implemented in the VASP 5.4 code. Valence electrons (Ag: 4d, 5s; F: 2s, 2p) were treated

explicitly, while standard VASP pseudopotentials (accounting for scalar relativistic effects) were used for the description of core electrons. The cutoff energy of the plane waves was set to 800 eV, a self-consistent-field convergence criterion to $10^{-6}$ eV, Gaussian smearing width to 0.05 eV, and the k-point mesh was set to $2\pi \times 0.03$ Å$^{-1}$. For band gap calculations the HSE06 hybrid functional was used.[33]

§ In this work we use the Heisenberg Hamiltonian in the form $H = -(½) \sum_{ij} J_{ij} s_i s_j$ with ($J_{ij} = J_{ji}$). Negative values of *J* mean AFM interactions. The ½ factor eliminates double counting.

§§ This underestimation might be a result of an error. The energies of the spin states reported in Table 4 of ref. 20 clearly indicate that $J_t'$ (labelled as $J_1$ in ref. 20) should range from –580 meV to –319 meV depending on the value of the Coulomb repulsion parameter (*U*) used by the authors. This contrast the values reported in their Table 5 which range from –37 meV to –20 meV (*cf.* ESI).

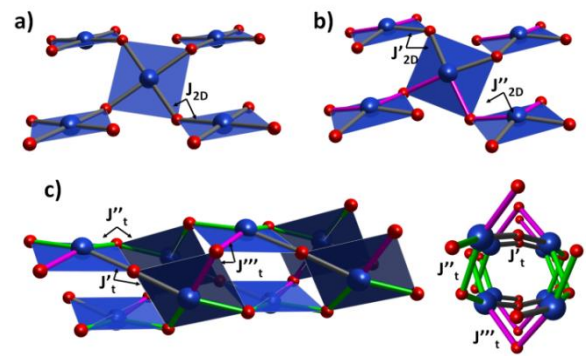

**Fig. 1** Main structural features in polymorphs of AgF$_2$: sheets in LP (a) and HP1 (b), as well as nanotubes in HP2 (c). Equivalent SE pathways are identified with colour bonds and the corresponding exchange constants are labelled.

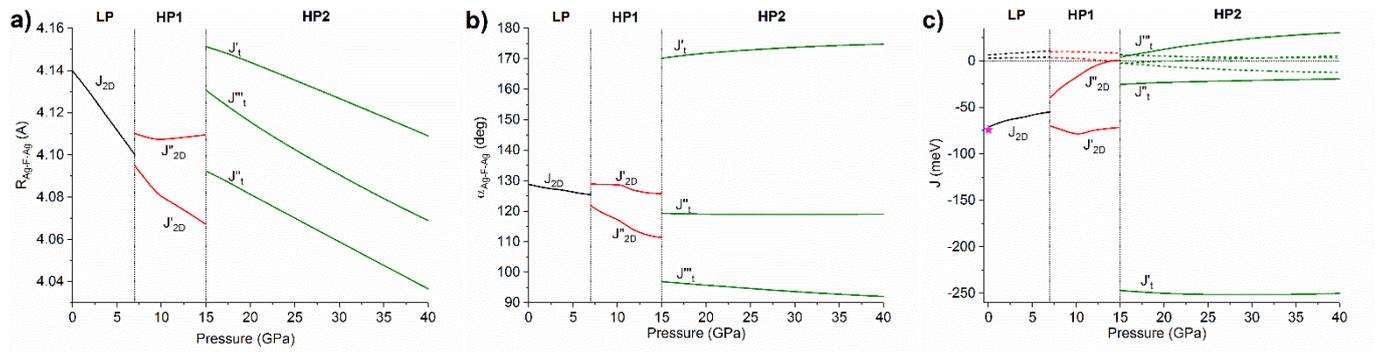

**Fig. 2** Pressure dependence of the sum of the Ag-F bond lengths along the SE routes (a), the Ag-F-Ag angle (b), and the values of the coupling constants along the SE routes (c). Black/red/green lines denote values for LP/HP1/HP2. Colour dashed lines in (c) show values of inter-sheet (LP, HP1) and inter-tube (HP2) coupling constants (see ESI). Vertical dashed lines enclose the region of stability of the respective $AgF_2$ polymorphs. The star in (c) indicates the experimental value of $J_{2D}$ for LP at ambient pressure (-70 meV).[4]

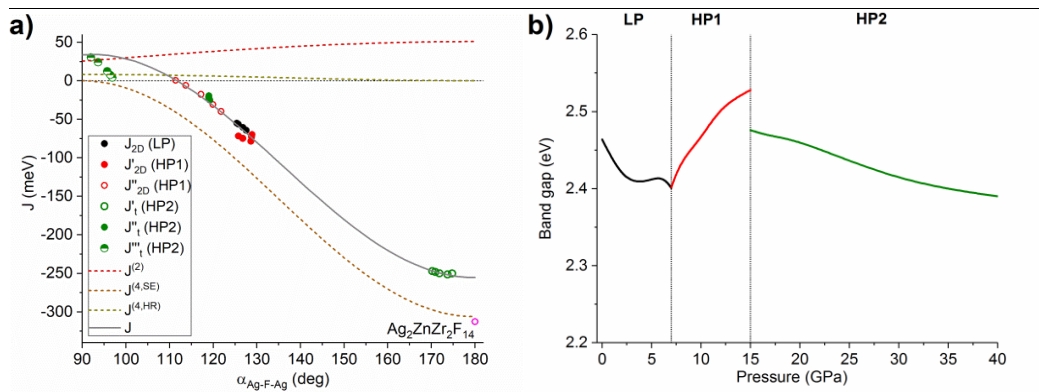

**Fig. 3** Panel (a) shows the dependence of the values of the coupling constants for AgF$_2$ polymorphs on the angle of the Ag-F-Ag bridge. The *J* value for dimers found in Ag$_2$ZnZr$_2$F$_{14}$ is also shown. The full line show the result of an analytical computation of *J* for a single Ag-F-Ag bridge and the dashed lines show the three contributions considered: a ferromagnetic contribution due to direct $pd$ exchange ($J^{(2)}$), the SE contribution ($J^{(4,SE)}$) and a ferromagnetic contribution due to Hund's exchange in F ($J^{(4,HR)}$). For details see text and ESI. Panel (b) shows the variation of the band gap for (LP – black; HP1 – red; HP2 – green line).

# Supplementary Information (ESI)

## Calculations details

In the present study we have used structures of $AgF_2$ polymorphs derived from DFT+U calculations performed in ref. 6. The supercells of the LP, HP1, and HP2 polymorphs used in this study are given, for selected pressures, at the end of this document.

## Magnetic Interactions

Our propose here is to illustrate with a simple analytical model the origin of the strong angular dependence found in the main text, and the sign and order of magnitude of the interaction in the different configurations. We consider an Ag-F-Ag bridge. The exchange interaction can be computed using perturbation theory in the $pd$ hopping matrix elements. We take the same setting as in Ref. 4 with one $d_{x^2-y^2}$ in each Ag and two $p$ orbitals in the bridging F: $p_\parallel$, oriented parallel to the Ag-Ag direction and $p_z$ which is oriented perpendicular to the Ag-Ag direction but parallel to the triangle formed by the Ag-F-Ag complex. Since most parameters are not known we take only a minimal set of parameters different from zero to illustrate the main trends. The dependence on the angle $\alpha$ of the bridge is determined by the hopping matrix elements $t_{\parallel d}$ between $p_\parallel$ and $d_{x^2-y^2}$ and $t_{zd}$ between $p_z$ and $d_{x^2-y^2}$ which can be parameterized in terms of Slater-Koster[25] integrals as $t_{\parallel d} = t_{pd} \sin(\alpha/2)$ and $t_{zd} = t_{pd} \cos(\alpha/2)$. Here $t_{pd}$ is the hopping matrix element for a straight bond. In addition there is a dependence of the hopping matrix element on the Ag-F distance which can be parameterized as $t_{pd} \sim 1/R^4$. Due to the small changes in bond length with pressure, the correction due to this effect is much less relevant and will be neglected in the present simplified treatment. Therefore, all the computations are done keeping the $t_{pd}$ the same for all bond lengths. The magnetic interaction is computed as the energy difference between the singlet and the triplet in the Ag-F-Ag system and reads,

$$J = J^{(2)} + J^{(4,SE)} + J^{(4,HR)},$$

where $J^{(2)} \propto t_{\parallel d}^2$ is a ferromagnetic contribution due to direct exchange, $K_{\parallel d} > 0$ between the $d_{x^2-y^2}$ and the $p_\parallel$ orbital,

$$J^{(2)} = t_{pd}^2 \sin^2\left(\frac{\alpha}{2}\right) \left[\frac{1}{(\Delta - K_{\parallel d})} - \frac{1}{(\Delta - K_{\parallel d})}\right],$$

$J^{(4,SE)} \propto \left(t_{\parallel d}^2 - t_{zd}^2\right)^2$ is the superexchange antiferromagnetic contribution,

$$J^{(4,SE)} = -t_{pd}^4 \cos^2\alpha \frac{1}{\Delta^2}\left[\frac{4}{U_d} + \frac{8}{(2\Delta + U_p)}\right],$$

and $J^{(4,HR)} \propto t_{\parallel d}^2 t_{zd}^2$ is a ferromagnetic contribution due to Hund's rule exchange interaction $J_H > 0$ on fluorine,

$$J^{(4,HR)} = t_{pd}^4 \sin^2\alpha \frac{8}{\Delta^2}\left[\frac{1}{(2\Delta + U_p - J_H)} - \frac{1}{(2\Delta + U_p)}\right].$$

Similar expressions were presented in the supplementary information to Ref. 4 except that $J^{(4,HR)}$ contribution was neglected. In Fig. 3a of the main text we compare these perturbative expressions with the results obtained with the full DFT solution. As a reference parameter set we took, $K_{\parallel d} = 0.07\text{eV}$, $J_H = 0.7\text{eV}$ for Hund's interaction among $p$ orbitals, $\Delta = 2.7\text{eV}$ for the charge transfer energy, $U_d = 9.4\text{eV}$ ( $U_p = 4\text{eV}$) for the Hubbard interaction on Ag (F) and $t_{pd} = 1.15\text{eV}$.

We see from Fig. 3a of the main paper that the superexchange contribution, $J^{(4,SE)}$, is in general dominant except close to $\alpha = 90°$ where it vanishes. Around this angle the sign of the magnetic interaction gets reversed due to the direct exchange contributions and in agreement with Goodenough-Kanamori-Anderson rules. $J^{(4,HR)}$ appears at fourth order so it is a much smaller contribution than $J^{(2)}$ and vanishes in the case of a straight bond where only $p_\parallel$ is relevant. For $J^{(2)}$ , we only took into account the direct exchange interaction with the $p_\parallel$ orbital. A more accurate treatment would consider the direct exchange with both orbitals and the angular dependence of the $K$'s matrix elements.

The parameters used are similar but not equal to the ones of Ref. 4 which were optimized only for zero pressure. There is considerable freedom for the choice of parameters. For example, practically an equally good fit of the DFT data is obtained with $U_d = 6\text{eV}$ and $t_{pd} = 1.1\text{eV}$. A more realistic computation would take into account also four-ring exchange processes that are expected to be important in covalent materials. Also, an accurate parameterization would require to take into account the effects of other orbitals which, however, would jeopardize our intention to get a simple understanding of the trends. With these caveats, we see that the general trends of the DFT computation are reproduced and the main microscopic matrix elements giving rise to the interactions are identified.

## Magnetic Topology

In order to discuss the possible magnetic ground state of the system it is useful to identify the motifs formed by the more relevant superexchange paths identified.

For the LP phase, layer directions are equivalent from a magnetic point of view so in a first approximation the system can be described by the two-dimensional Heisenberg antiferromagnet within layers (Supplementary Fig. 1a) with weak coupling among layers. The ground state has robust 2D antiferromagnetic order inside the layers which becomes long-range three-dimensional order below the Néel temperature. Neglecting spin-orbit coupling the magnetic excitation spectrum is gapless due to acoustic spin-wave modes.

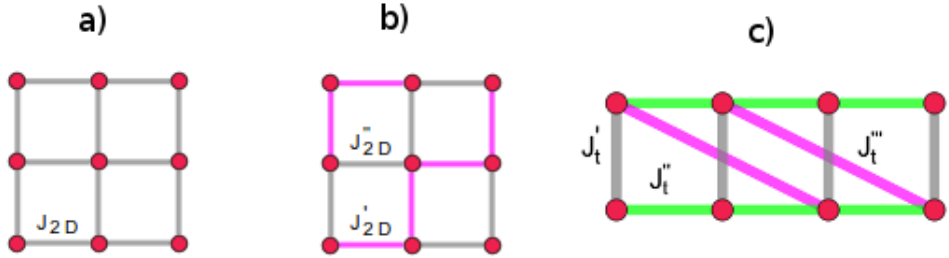

**Supplementary Figure 1**: Magnetic lattices corresponding to the AgF$_2$ polymorphs shown in Fig. 1 of the main text. The structure can be seen as formed by sheets in LP phase(a) zig-zag chains in HP1 (b), and ladders in HP2 (c).

For the HP1 phase, two non-equivalent paths appear within the layers forming a zig-zag pattern (Supplementary Fig. 1b). Lowering the temperature the system is expected to show 1D quasi-long range order along the $J'_{2D}$ zig-zag chains at high temperatures which becomes more 2D as the temperature is lower and finally 3D long-range antiferromagnetic order below the Néel temperature. Close to the critical pressure to the HP2 phase the $J''_{2D}$ coupling vanishes and the system becomes effectively 1D at all temperatures. The ground state is gapless unless a dimerization and a further lowering of symmetry (not considered here) occurs.

For the HP2 phase, the $J''_t$ chains can be seen as the legs of a ladder. We conventionally take $J'_t$ as being the rungs of the ladder by a fictitious displacement of one leg respect to the other. With this setting $J'''_t$ forms "diagonal" rungs (Fig. 3c). The ground state consists of 0D singlets defined on the rungs of the ladder[22] and remains gapped up to zero temperature unless weaker couplings among ladders drive 3D antiferromagnetic order at low temperature. Given the large difference in values of the interactions we think the 0D-singlet ground state scenario is more likely.

## Effect of spin-orbit coupling

The effect of spin orbit coupling was analyzed using the GGA+U method (U=5eV) as implemented in VASP. We checked that the ordering of the phases is not altered respect to SCAN meta-GGA without spin-orbit coupling.

Supplementary Figure 1 shows the magnetic structure of the LP polymorph. The figure setting reproduces Fig. 3 of Ref. 2. In excellent agreement with this reference we find that magnetic moments are approximately directed in the crystallographic **a** direction.

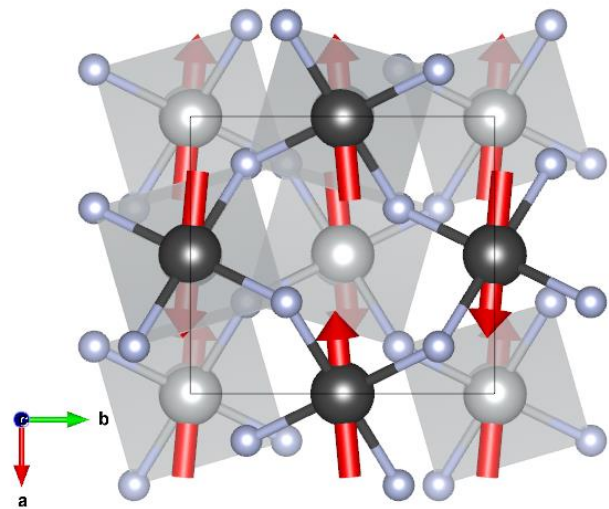

**Supplementary Figure 2:** Magnetic structure of LP polymorph to be compared with the experimental result in Ref 2. Dark Ag atoms are at z=1/2 while light gray are at z=0. There is a trivial inversion of the magnetic moments in the *ab* plane respect to Fig. 3 of Ref. 2 related to the choice of origin of the unit cell.

## $Ag_2ZnZr_2F_{14}$ reference system

$Ag_2ZnZr_2F_{14}$ was chosen as a reference system due to the presence of planar $[AgF_4]^{2-}$ units connected via a single F atom bridge forming $Ag_2F_7^{3-}$ dimers with straight Ag–F–Ag bridges in its crystal structure (Supplementary Figure 3).[23]

We have evaluated the superexchange coupling constants within the dimers (*J*, Fig. 3a), as well as along two inter-dimer routes ($J_{i1}$, Fig 3b, $J_{i2}$, Fig 3c). The Ag⋯Ag distances along the *J*, $J_{i1}$, and $J_{i2}$ routes are 4.03 Å, 3.97 Å, and 5.60 Å, respectively. Apart from these contacts there are no other Ag⋯Ag distances shorter than 5.8 Å.

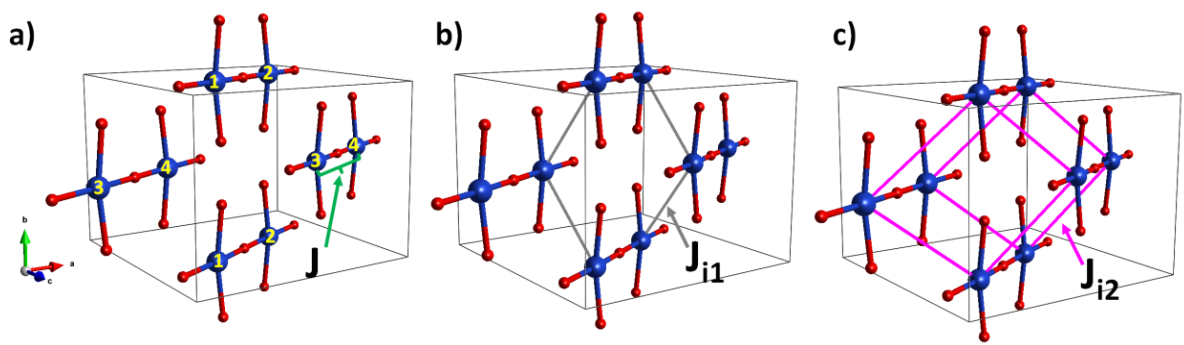

**Supplementary Figure 3.** The Ag/F bonding framework of $Ag_2ZnZr_2F_{14}$ (a) featuring the main intradimer superexchange pathway, as measured by J, together with the depiction of the inter-dimer $J_{i1}$ (b) and $J_{i2}$ (c) superexchange routes.

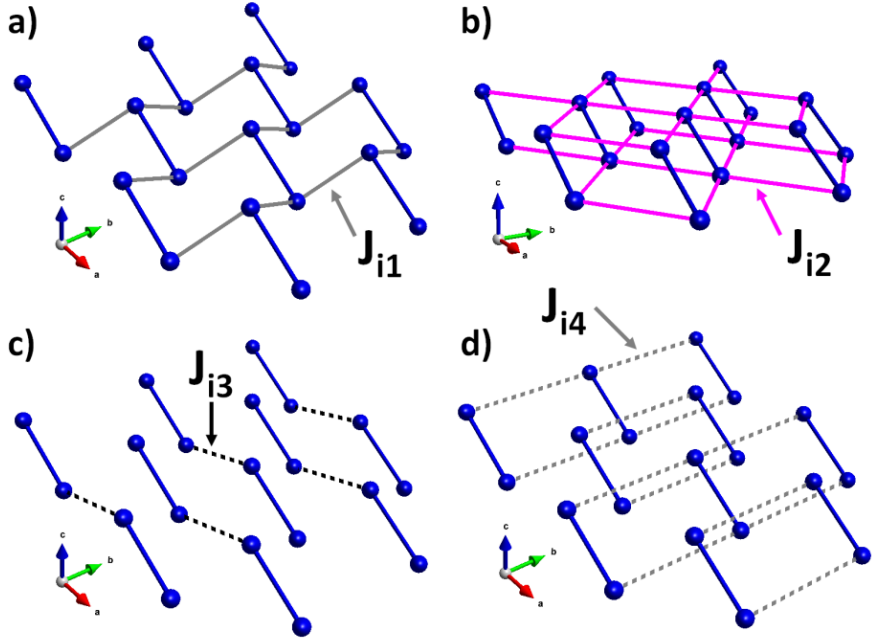

**Supplementary Figure 4.** Inter-dimer coupling constants in $Ag_2ZnZr_2F_{14}$.

The $J_{i1}$ and $J$ routes form a honeycomb-like lattice (Fig. 4a), while $J_{i2}$ connects the dimers into a square lattice (Fig. 4b). The SE routes not taken into account in this analysis are $J_{i3}$ (Fig 4c, Ag⋯Ag distances of 5.82 Å) which connect the dimers into chains, and $J_{i4}$ (Fig 4d, Ag⋯Ag distances of 6.64 Å) through which ladders are formed. As shown below already the $J_{i1}$ and $J_{i2}$ constants are small (absolute values smaller than 0.5 meV), and one should expect $J_{i3}$ and $J_{i4}$ to be negligibly small.

**Supplementary Table 1.** Magnetic states of $Ag_2ZnZr_2F_{14}$ and their energy. Spin up/down sites are indicated with a +/− sign, site labelling follows that of Supplementary Figure 3; $E_{nm}$ denotes the part of the total-energy of the system which is independent of the spin state.

| Site: | 1 | 2 | 3 | 4 | Energy per f.u. | Energy (eV per f.u.) |
|---|---|---|---|---|---|---|
| F1 | + | + | + | + | $-0.125J - 0.25J_{i1} - 0.5J_{i2} + E_{nm}$ | −120.8945 |
| F2 | + | + | − | − | $-0.125J + 0.25J_{i1} + 0.5J_{i2} + E_{nm}$ | −120.8945 |
| A1 | + | − | + | − | $0.125J + 0.25J_{i1} - 0.5J_{i2} + E_{nm}$ | −120.9725 |
| A2 | + | − | − | + | $0.125J - 0.25J_{i1} + 0.5J_{i2} + E_{nm}$ | −120.9729 |
| | | | | | **Super-exchange constants** | **Super-exchange constants (meV)** |
| | | | | | $J = 2E_{A1} + 2E_{A2} - 2E_{F1} - 2E_{F2}$ | $J = -312.7$ |
| | | | | | $J_{i1} = E_{A1} - E_{A2} - E_{F1} + E_{F2}$ | $J_{i1} = 0.4$ |
| | | | | | $J_{i2} = 0.5(-E_{A1} + E_{A2} - E_{F1} + E_{F2})$ | $J_{i2} = -0.2$ |

The coupling constants were extracted with the broken symmetry method through equations given in Supplementary Table 1. The same method was for applied in an earlier study of the SE coupling in $Ag_2ZnZr_2F_{14}$.[20] They used spin states with the following energies:

$E_{AF1} = (+4J_1-8J_2+8J_3)(1/4)$

$E_{AF2} = (-4J_1+8J_2)(1/4)$

$E_{AF3} = (+4J_1+8J_2)(1/4)$

$E_{AF3} = (+4J_1-8J_2-8J_3)(1/4)$

where $J_1$ is the intra-dimer coupling constant and $J_2$ and $J_3$ are inter-dimer couplings. From the above one can see that:

$J_1 = 1/2(E_{AF3} - E_{AF2})$

From Table 4 of ref. 20 one can easily see that the AF2 states lies 1174.81 to 638.6 meV higher than AF3 for the $U$ parameter in the DFT+U method equal from 2 to 6 eV. This gives intra-dimer coupling constants in the range from –587.4 to –319.3 meV. The latter value, which corresponds to a more realistic $U$ value in the case of silver(II) compounds, is close to that computed in the present work.

## Coupling constant calculations for LP, HP1, and HP2

For the LP polymorph (Fig. 5a) we have evaluated the intra-sheet coupling constant ($J_{2D}$) and two inter-sheet ones ($J_{i1}$, $J_{i2}$) as shown in Fig. 5b.

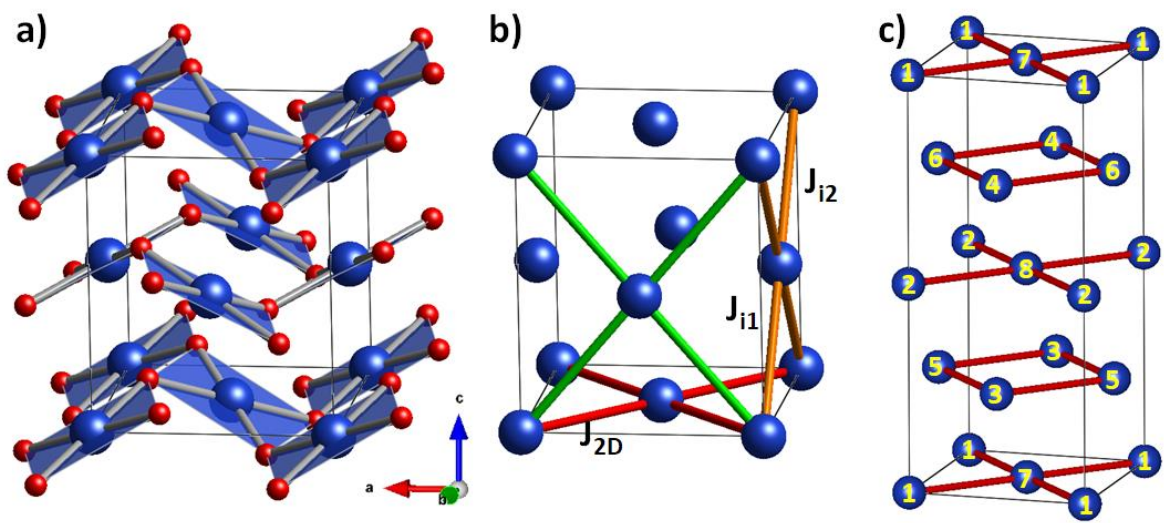

**Supplementary Figure 5.** The Ag/F bonding framework of LP-$AgF_2$ (a), together with the depiction of the intra-sheet coupling constant ($J_{2D}$), as well as the two inter-sheet SE routes (b). The 1x1x2 supercell used in the calculations is depicted in (c).

**Supplementary Table 2.** Magnetic states of LP-AgF$_2$ and their energy. Spin up/down sites are indicated with a +/− sign, site labelling follows that of Supplementary Figure 5c; E$_{nm}$ denotes the part of the total-energy of the system which is independent of the spin state.

| Site: | 1 | 2 | 3 | 4 | 5 | 6 | 7 | 8 | Energy per f.u. |
|---|---|---|---|---|---|---|---|---|---|
| F4 | + | + | + | + | − | − | + | + | $E_{nm}$ |
| A1 | + | + | + | + | − | − | − | − | $0.5J_{2D} - 0.5J_{i1} + 0.5J_{i2} + E_{nm}$ |
| A2 | + | + | − | − | − | − | + | + | $-0.5J_{2D} + 0.5J_{i1} + 0.5J_{i2} + E_{nm}$ |
| A4 | + | + | − | − | + | + | − | − | $0.5J_{2D} + 0.5J_{i1} - 0.5J_{i2} + E_{nm}$ |
| **Super-exchange constants** ||||||||||
| $J_{2D} = E_{A1} + E_{A4} - 2E_{F4}$ ||||| $J_{i1} = E_{A2} + E_{A4} - 2E_{F4}$ ||| $J_{i2} = E_{A1} + E_{A2} - 2E_{F4}$ ||

In case of the HP1 polymorph the basic cell is analogous to that of LP (Fig. 5a/5c), but due to the alternation in the Ag-F-Ag angle there are two intra-sheet couplings ($J'_{2D}$ and $J''_{2D}$). The inter-sheet SE also split into pairs ($J'_{i1}/J''_{i1}$ and $J'_{i2}/J''_{i2}$), but given their small value we have evaluated only their mean, that is $J^{mean}_{i1} = ½( J'_{i1} + J''_{i1})$ and $J^{mean}_{i2} = ½( J'_{i2} + J''_{i2})$. The corresponding equations are given in Table 3.

**Supplementary Table 3.** Magnetic states of HP1-AgF$_2$ and their energy. Spin up/down sites are indicated with a +/− sign, site labelling follows that of Supplementary Figure 5c, except for the F7 state which is depicted in Supplementary Figure 6; E$_{nm}$ denotes the part of the total-energy of the system which is independent of the spin state.

| Site: | 1 | 2 | 3 | 4 | 5 | 6 | 7 | 8 | Energy per f.u. |
|---|---|---|---|---|---|---|---|---|---|
| F3 | + | + | − | + | − | + | + | + | $-0.25J'_{2D} - 0.25J''_{2D} + E_{nm}$ |
| F4 | + | + | + | + | − | − | + | + | $E_{nm}$ |
| F7 | | | see Fig. 6 | | | | | | $0.25J'_{2D} - 0.25J''_{2D} + E_{nm}$ |
| A1 | + | + | + | + | − | − | − | − | $0.25J'_{2D} + 0.25J''_{2D} - 0.5J^{mean}_{i1} + 0.5J^{mean}_{i2} + E_{nm}$ |
| A2 | + | + | − | − | − | − | + | + | $-0.25J'_{2D} - 0.25J''_{2D} + 0.5J^{mean}_{i1} + 0.5J^{mean}_{i2} + E_{nm}$ |
| A3 | + | − | + | + | − | − | − | + | $0.25J'_{2D} + 0.25J''_{2D} + E_{nm}$ |
| A4 | + | + | − | − | + | + | − | − | $0.25J'_{2D} + 0.25J''_{2D} + 0.5J^{mean}_{i1} - 0.5J^{mean}_{i2} + E_{nm}$ |
| **Super-exchange constants** ||||||||||
| $J'_{2D} = 2E_{A3} - 2E_{F7}$ $J''_{2D} = 2E_{F7} - 2E_{F3}$ ||||| $J^{mean}_{i1} = E_{A2} + E_{A4} - 2E_{F4}$ ||| $J^{mean}_{i2} = E_{A1} + E_{A2} - 2E_{F4}$ ||

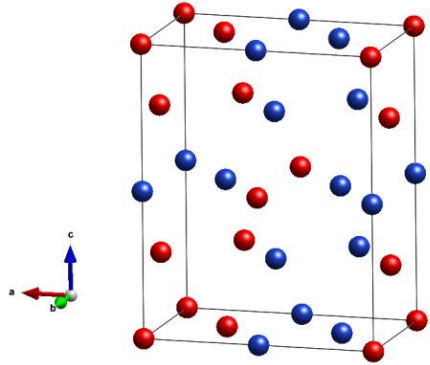

**Supplementary Figure 6.** The 2x1x2 supercell of HP1 used in the calculations of the F7 magnetic state (see Table 3). Spin up/down sites are marked with blue/red balls.

For the HP2 structure apart from the intra-tube SE routes (Supplementary Figure 7a) we considered three inter-tube couplings ($J_{i1-i3}$, Supplementary Figure 7b-d).

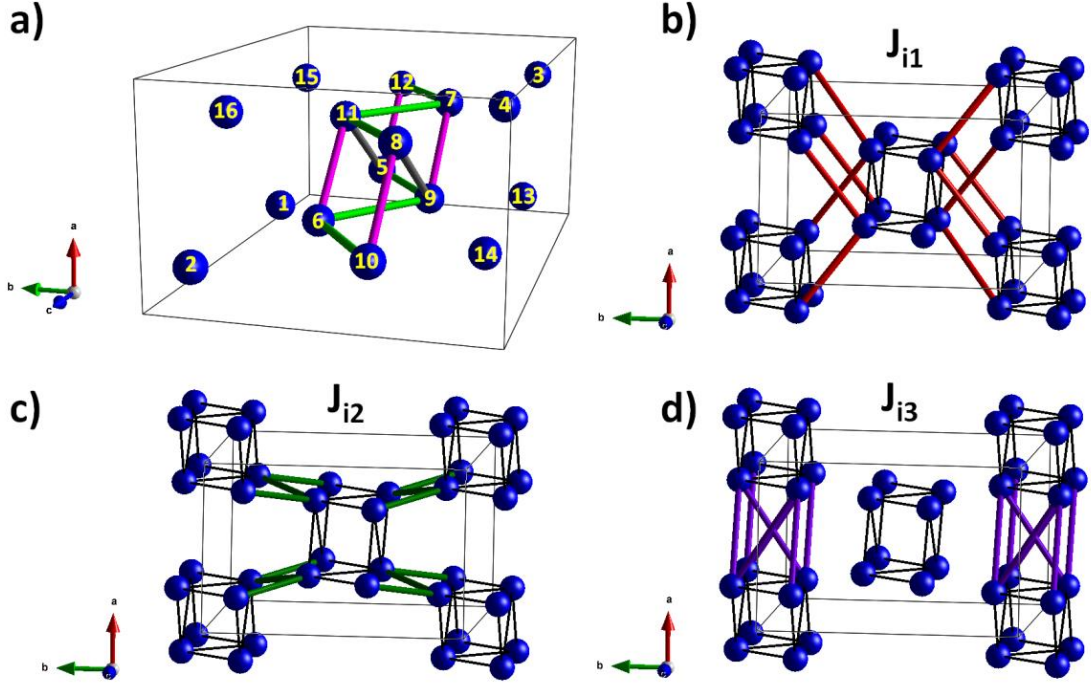

**Supplementary Figure 7.** The 1x1x2 supercell of HP2 used in the calculations depicting the intra-tube SE routes (a), as well as the inter-tube ones (b-d).

**Supplementary Table 4.** Magnetic states of HP2-AgF$_2$ and their energy. Spin up/down sites are indicated with a +/− sign, site labelling follows that of Supplementary Figure 7a.

| Site: | 1 | 2 | 3 | 4 | 5 | 6 | 7 | 8 | 9 | 10 | 11 | 12 | 13 | 14 | 15 | 16 |
|---|---|---|---|---|---|---|---|---|---|---|---|---|---|---|---|---|
| **A1** | + | + | − | − | + | + | − | − | + | + | − | − | + | + | − | − |
| **A2** | + | + | + | + | + | + | + | + | − | − | − | − | − | − | − | − |
| **A3** | + | − | − | + | − | + | + | − | + | − | − | + | − | + | + | − |
| **F1** | + | + | + | + | + | + | + | + | + | + | + | + | + | + | + | + |
| **F2** | + | + | + | + | − | − | − | − | − | − | − | − | − | − | − | − |
| **F3** | + | + | − | − | − | − | + | + | + | + | − | − | − | − | + | + |
| **F4** | + | + | − | − | − | − | + | + | + | + | − | − | − | − | + | + |

**Super-exchange constants**

$$J'_t = -1.6E_{F4} - 1.2E_{F3} - 1.6E_{F2} - 0.8E_{F1} + 4E_{A3} + 4E_{A2} + 0.8E_{A1}$$
$$J''_t = -1.0666E_{F4} + 0.5333E_{F3} + 1.6E_{F2} - 0.5333E_{F1} + 0.9333E_{A2} - 1.4666E_{A1}$$
$$J'''_t = 1.6E_{F4} - 2.8E_{F3} + 1.6E_{F2} + 0.8E_{F1} - 4E_{A3} + 3.6E_{A2} - 0.8E_{A1}$$
$$J_{i1} = 2.1333E_{F4} + 0.9333E_{F3} - 3.2E_{F2} - 0.9333E_{F1} + 0.1333E_{A2} + 0.9333E_{A1}$$
$$J_{i2} = 1.6E_{F4} + 0.8E_{F3} + 1.6E_{F2} - 1.2E_{F1} - 0.4E_{A2} - 0.8E_{A1}$$
$$J_{i3} = 0.1333E_{F4} - 0.0666E_{F3} + 0.8E_{F2} + 0.0666E_{F1} + 0.1333E_{A2} - 0.0666E_{A1}$$

## Structures in VASP format

Below we give the structural information (in VASP format) for the supercells of LP (at 4 GPa), HP1 (at 12 GPa), and HP2 (at 30 GPa) which were used in the calculations. The lattice constants and the fractional coordinates are taken unchanged from our previous work (Ref.6 in the main paper) – in that work they have been taken from DFT+U calculations which were then used to propose various polymorphic forms to be tested in Rietveld fits to experimental data.

```
LP @ 4 GPa
1.0
     4.9556121826     0.0000000000     0.0000000000
     0.0000000000     5.4461851120     0.0000000000
     0.0000000000     0.0000000000    11.3111810684
  Ag   F
   8  16
Direct
   0.000000000      0.000000000      0.000000000
   0.000000000      0.000000000      0.500000000
   0.500000000      0.000000000      0.250000000
   0.500000000      0.000000000      0.750000000
   0.000000000      0.500000000      0.250000000
   0.000000000      0.500000000      0.750000000
   0.500000000      0.500000000      0.000000000
   0.500000000      0.500000000      0.500000000
   0.169777229      0.315485209      0.433581322
   0.169777229      0.315485209      0.933581352
   0.830222785      0.684514761      0.066418663
   0.830222785      0.684514761      0.566418648
   0.669777215      0.184514791      0.066418663
   0.669777215      0.184514791      0.566418648
   0.330222785      0.815485239      0.433581322
   0.330222785      0.815485239      0.933581352
   0.330222785      0.684514761      0.183581337
   0.330222785      0.684514761      0.683581352
   0.669777215      0.315485209      0.316418678
   0.669777215      0.315485209      0.816418648
   0.830222785      0.815485239      0.316418678
   0.830222785      0.815485239      0.816418648
   0.169777229      0.184514791      0.183581337
   0.169777229      0.184514791      0.683581352

HP1 @ 12 GPa
1.0
     4.4479866028     0.0000000000     0.0000000000
     0.0000000000     5.5061345100     0.0000000000
     0.0000000000     0.0000000000    11.1615753174
```

```
  Ag   F
   8   16
Direct
  0.999999046    0.999995649    0.000001085
  0.000000000    0.999995649    0.500001073
  0.463440925    0.086044364    0.250001073
  0.463440925    0.086044364    0.750001073
  0.999999046    0.586044371    0.250001073
  0.000000000    0.586044371    0.750001073
  0.463440925    0.499995649    0.000001085
  0.463440925    0.499995649    0.500001073
  0.344793469    0.767811835    0.166536614
  0.344793469    0.767811835    0.666536629
  0.608406901    0.422753751    0.299424320
  0.608406901    0.422753751    0.799424291
  0.855033040    0.922753751    0.299424320
  0.855033040    0.922753751    0.799424291
  0.118646532    0.267811835    0.166536614
  0.118646532    0.267811835    0.666536629
  0.118646532    0.318228185    0.416536599
  0.118646532    0.318228185    0.916536570
  0.855033040    0.663286269    0.049424332
  0.855033040    0.663286269    0.549424350
  0.608406901    0.163286254    0.049424332
  0.608406901    0.163286254    0.549424350
  0.344793469    0.818228126    0.416536599
  0.344793469    0.818228126    0.916536570
```

---

```
HP2 @ 30 GPa
1.0
     5.0722217560     0.0000000000     0.0000000000
     0.0000000000     7.8814549446     0.0000000000
     0.0000000000     0.0000000000    11.4406347275
  Ag   F
  16   32
Direct
  0.203389764    0.872389853    0.469119519
  0.203389764    0.872389853    0.969119549
  0.796610236    0.127610177    0.030880490
  0.796610236    0.127610177    0.530880511
  0.296610236    0.627610147    0.219119504
  0.296610236    0.627610147    0.719119489
  0.703389764    0.372389853    0.280880481
  0.703389764    0.372389853    0.780880451
  0.296610236    0.372389853    0.469119519
  0.296610236    0.372389853    0.969119549
```

| | | |
|---|---|---|
| 0.703389764 | 0.627610147 | 0.030880490 |
| 0.703389764 | 0.627610147 | 0.530880511 |
| 0.203389764 | 0.127610177 | 0.219119504 |
| 0.203389764 | 0.127610177 | 0.719119489 |
| 0.796610236 | 0.872389853 | 0.280880481 |
| 0.796610236 | 0.872389853 | 0.780880451 |
| 0.500000000 | 0.358051956 | 0.125000000 |
| 0.500000000 | 0.358051956 | 0.625000000 |
| 0.500000000 | 0.805217743 | 0.125000000 |
| 0.500000000 | 0.805217743 | 0.625000000 |
| 0.179253072 | 0.413714439 | 0.300836891 |
| 0.179253072 | 0.413714439 | 0.800836921 |
| 0.320746928 | 0.913714468 | 0.300836891 |
| 0.320746928 | 0.913714468 | 0.800836921 |
| 0.179253072 | 0.586285532 | 0.050836906 |
| 0.179253072 | 0.586285532 | 0.550836921 |
| 0.320746928 | 0.086285546 | 0.050836906 |
| 0.320746928 | 0.086285546 | 0.550836921 |
| 0.500000000 | 0.641948044 | 0.375000000 |
| 0.500000000 | 0.641948044 | 0.875000000 |
| 0.500000000 | 0.194782287 | 0.375000000 |
| 0.500000000 | 0.194782287 | 0.875000000 |
| 0.000000000 | 0.694782257 | 0.375000000 |
| 0.000000000 | 0.694782257 | 0.875000000 |
| 0.000000000 | 0.141948074 | 0.375000000 |
| 0.000000000 | 0.141948074 | 0.875000000 |
| 0.820746899 | 0.586285532 | 0.199163094 |
| 0.820746899 | 0.586285532 | 0.699163079 |
| 0.679253101 | 0.086285546 | 0.199163094 |
| 0.679253101 | 0.086285546 | 0.699163079 |
| 0.820746899 | 0.413714439 | 0.449163109 |
| 0.820746899 | 0.413714439 | 0.949163079 |
| 0.679253101 | 0.913714468 | 0.449163109 |
| 0.679253101 | 0.913714468 | 0.949163079 |
| 0.000000000 | 0.305217743 | 0.125000000 |
| 0.000000000 | 0.305217743 | 0.625000000 |
| 0.000000000 | 0.858051956 | 0.125000000 |
| 0.000000000 | 0.858051956 | 0.625000000 |